%% file: sosp.tex
\documentclass[sigplan,10pt,nonacm,preprint,screen]{acmart}

\renewcommand\footnotetextcopyrightpermission[1]{}
\usepackage{amsmath,amssymb,amsfonts}
\usepackage{algorithmic}
\usepackage{graphicx}
\usepackage{textcomp}
\usepackage{xcolor}
\usepackage{caption}
\usepackage{subcaption}
\usepackage{bbding}

\usepackage{ulem}
\definecolor{Color1}{RGB}{200,36,35}
\definecolor{Color2}{RGB}{40,120,181}
\usepackage{algorithmic}
\usepackage{graphicx}
\usepackage{textcomp}
\usepackage{color}
\usepackage{url}
\usepackage{enumitem}
\usepackage{multirow}
\usepackage[final]{pdfpages}

\usepackage[ruled, vlined, linesnumbered]{algorithm2e}
\SetAlCapNameFnt{\scriptsize}
\SetAlCapFnt{\small}
\usepackage{hyperref}
\usepackage{listings}

\definecolor{keywordcolor}{RGB}{0,0,187}        
\definecolor{commentcolor}{RGB}{0,128,0}        
\definecolor{stringcolor}{RGB}{187,0,0}         
\definecolor{numbercolor}{RGB}{0,153,153}       
\definecolor{preprocesscolor}{RGB}{128,0,128}   

\usepackage{xspace}

\usepackage{soul}

\newcommand*{\eg}{e.g., }

\usepackage{stackengine}
\usepackage{booktabs}

\newcommand{\mypara}[1]{\vspace{8pt}\noindent\textit{\textbf{#1} }}

\usepackage{fbox}
\usepackage[most]{tcolorbox}

\soulregister{\cite}7 
\soulregister{\citep}7 
\soulregister{\citet}7 
\soulregister{\ref}7 
\soulregister{\pageref}7 
\soulregister{\tool}7
\soulregister{\eg}7

\usepackage{colortbl}
\usepackage{pgfplots}
\usepackage{microtype}
\usepgfplotslibrary{dateplot}
\pgfplotsset{compat=newest}
\colorlet{shadecolor}{gray!40}

\usepackage{balance}

\usepackage{tikz}
\usetikzlibrary{shapes.geometric,arrows,shapes,shadows,patterns}

\usepackage{pifont}
\usepackage{flushend}

\definecolor{mygreen}{HTML}{02818a}

\newcommand{\tool}{\textsc{DebugHarness}}

\newcommand{\dingnum}[1]{%
  \ifcase#1
    \or \ding{182} %
    \or \ding{183} %
    \or \ding{184} %
    \or \ding{185} %
    \or \ding{186} %
    \or \ding{187} %
    \or \ding{188} %
    \or \ding{189} %
    \or \ding{190} %
    \or \ding{191} %
  \fi
}

\newcommand{\keepnotes}{true} 

\ifthenelse{\equal{\keepnotes}{true}}{
  \newcommand{\mytodogreen}[1]{\textcolor{mygreen}{\ding{46}~{\sf}~#1}}
\newcommand{\yyb}[1]{\\newcommand{\yyb}[1]{\textcolor{mygreen}{[yyb: #1]}}{mygreen}{[yyb: #1]}}
\newcommand{\sml}[1]{\textcolor{blue}{[sml: #1]}}
}{
  \newcommand{\mytodogreen}[1]{}
  \newcommand{\yyb}[1]{}
  \newcommand{\sml}[1]{}
}

\newtcbox{\mybox}[1][darkgray]
  {on line, arc = 0pt, outer arc = 0pt,
    colback = #1!10!white, colframe = #1!50!black,
    boxsep = 0pt, left = 1pt, right = 1pt, top = 2pt, bottom = 2pt,
    boxrule = 0pt, bottomrule = 0pt, toprule = 0pt}

\newcounter{finding}
\newtcolorbox[use counter=finding]{findingbox}{
  colback=green!5!white,      
  colframe=green!40!black,    
  fonttitle=\bfseries,        
  coltitle=white,             
  title=Finding~\thefinding,  
  boxrule=1pt,                
  arc=3mm,                    
  top=3mm, bottom=3mm,        
  left=4mm, right=4mm,        
  enhanced,
  attach boxed title to top left={xshift=6mm,yshift=-3mm},
  boxed title style={size=small,colback=green!60!black,arc=2mm},
  drop shadow={opacity=0.3,shadow xshift=1mm,shadow yshift=-1mm}
}

\begin{document}

\title{\textsc{DebugHarness}: Emulating Human Dynamic Debugging for Autonomous Program Repair}

\author{Maolin Sun}
\orcid{0000-0001-5617-2205}
\affiliation{%
  \institution{State Key Laboratory for Novel Software Technology, Nanjing University}
  \city{Nanjing 210023}
  \country{China}
}
\email{merlin@smail.nju.edu.cn}

\author{Yibiao Yang}
\orcid{0000-0003-1153-2013}
\affiliation{%
  \institution{State Key Laboratory for Novel Software Technology, Nanjing University}
  \city{Nanjing 210023}
  \country{China}
}
\email{yangyibiao@nju.edu.cn}
\authornote{Corresponding author.}

\author{Xuanling Liu}
\affiliation{%
  \institution{State Key Laboratory for Novel Software Technology, Nanjing University}
  \city{Nanjing 210023}
  \country{China}
}
\email{522025330062@smail.nju.edu.cn}

\author{Yuming Zhou}
\orcid{0000-0002-4645-2526}
\affiliation{%
  \institution{State Key Laboratory for Novel Software Technology, Nanjing University}
  \city{Nanjing 210023}
  \country{China}
}
\email{zhouyuming@nju.edu.cn}

\author{Baowen Xu}
\orcid{0000-0001-7743-1296}
\affiliation{%
  \institution{State Key Laboratory for Novel Software Technology, Nanjing University}
  \city{Nanjing 210023}
  \country{China}
}
\email{bwxu@nju.edu.cn}

\begin{abstract}
\input{content/abstract.tex}
\end{abstract}



\settopmatter{printfolios=true}
\maketitle

\input{figure/plot.tex}

\section{Introduction}
\label{sec:intro}
\input{content/introduction.tex}

\section{Background and Motivation}
\label{sec:motivation}

\input{content/motivation.tex}


\section{Design}
\label{sec:design}
\input{content/approach.tex}

\section{Experimental Evaluation}
\label{sec:evaluation}
\input{content/evaluation.tex}

\section{Discussion}
\label{sec:discussion}
\input{content/discussion.tex}

\section{Related Work}
\label{sec:related-work}
\input{content/relatedworks.tex}

\section{Conclusion}
\label{sec:conclusion}
\input{content/conclusion.tex}

\balance
\bibliographystyle{ACM-Reference-Format}
\normalem
\bibliography{full}

\end{document}

%% file: content/abstract.tex
Patching severe security flaws in complex software remains a major challenge. While automated tools like fuzzers efficiently discover bugs, fixing deep-rooted low-level faults (e.g., use-after-free and memory corruption) still requires labor-intensive manual analysis by experts. Emerging Large Language Model (LLM) agents attempt to automate this pipeline, but they typically treat bug fixing as a purely static code-generation task. Relying solely on static artifacts, these methods miss the dynamic execution context strictly necessary for diagnosing intricate memory safety violations.

To overcome these limitations, we introduce \tool, an autonomous LLM-powered debugging agent harness that resolves complex vulnerabilities by emulating the interactive debugging practices of human systems engineers. Instead of merely examining static code, \tool\ actively queries the live runtime environment. Driven by a reproducible crash, it utilizes a pattern-guided investigation strategy to formulate hypotheses, interactively probes program memory states and execution paths, and synthesizes patches via a closed-loop validation cycle.

We evaluate \tool\ on SEC-bench, a rigorous dataset of real-world C/C++ security vulnerabilities. \tool\ successfully patches approximately 90\% of the evaluated bugs. This yields a relative improvement of over 30\% compared to state-of-the-art baselines, demonstrating that dynamic debugging significantly enhances LLM diagnostic capabilities. Overall, \tool\ establishes a novel paradigm for automated program repair, bridging the gap between static LLM reasoning and the dynamic intricacies of low-level systems programming.

%% file: figure/plot.tex
\newcommand{\commitsFilesZ}{
    \pgfplotstableread[row sep=\\,col sep=&]{
        interval & count \\
        1&185\\
        2&81\\
        3&46\\
        4&19\\
        5&22\\
        6&7\\
        7&4\\
        8&5\\
        9&1\\
        11&3\\
        13&1\\
        17&1\\
        55&1\\
        65&1\\
        }\mydata    
\begin{tikzpicture}
    \tikzset{every node}=[font=\tiny\sffamily]
    \begin{axis}[
        ybar,
        bar width=.20cm,
        width=.82\textwidth,
        height=4.0cm,
        legend style={at={(0.5,1)},
        anchor=north,legend columns=-1},
        symbolic x coords={1,2,3,4,5,6,7,8,9,11,13,17,55,65},
        xtick pos=left,
        ytick pos=left,
        xtick=data,
        enlarge x limits={abs=0.4cm},
        nodes near coords,
        nodes near coords align={vertical},
        ymin=0,ymax=250,
        ylabel={\#Commits},
        xlabel={\#File changes},
        label style={font=\footnotesize},
        ]
        \addplot[fill=black!80] table[x=interval,y=count]{\mydata};
    \end{axis}
\end{tikzpicture}
}

\newcommand{\commitsFilesCVC}{
    \pgfplotstableread[row sep=\\,col sep=&]{
        interval & count \\
        1&59\\
        2&24\\
        3&6\\
        4&6\\
        5&3\\
        6&1\\
        13&1\\
        18&1\\
        }\mydata
    \begin{tikzpicture}
        \tikzset{every node}=[font=\tiny\sffamily]
        \begin{axis}[
            ybar,
            bar width=.20cm,
            width=1\textwidth,
            height=3cm,
            legend style={at={(0.5,1)},
            anchor=north,legend columns=-1},
            symbolic x coords={1,2,3,4,5,6,13,18},
            xtick pos=left,
            ytick pos=left,
            xtick=data,
            enlarge x limits={abs=0.4cm},
            nodes near coords,
            nodes near coords align={vertical},
            ymin=0,ymax=82,
            ylabel={\#Commits},
            xlabel={\#File changes},
            label style={font=\footnotesize},
            ]
            \addplot[fill=white] table[x=interval,y=count]{\mydata};
        \end{axis}
    \end{tikzpicture}
}

\newcommand{\releaseZ}{
    \pgfplotstableread[row sep=\\,col sep=&]{
        interval & count \\
        4.8.1&3\\
        4.8.3&6\\
        4.8.4&7\\
        4.8.5&10\\
        4.8.6&6\\
        4.8.7&4\\
        4.8.8&3\\
        4.8.9&3\\
        4.8.10&4\\
        4.8.11&8\\
        4.8.12&5\\
        4.8.13&5\\
        4.8.14&5\\
        4.8.15&5\\
        4.8.16&5\\
        4.8.17&5\\
        4.9&6\\
        4.10&6\\
        4.11.0&6\\
        trunk&36\\
        }\mydata
    \begin{tikzpicture}
        \tikzset{every node}=[font=\scriptsize\sffamily]
        \begin{axis}[
            ybar,
            bar width=.2cm,
            width=0.5\textwidth,
            height=4.8cm,
            xticklabel style={rotate=40},
            legend style={at={(0.5,1)},
                anchor=north,legend columns=-1},
            symbolic x coords={4.8.1,4.8.3,4.8.4,4.8.5,4.8.6,4.8.7,4.8.8,4.8.9,4.8.10,4.8.11,4.8.12,4.8.13,4.8.14,4.8.15,4.8.16,4.8.17,4.9,4.10,4.11.0,trunk},
            xtick pos=left,
            ytick=\empty,
            xtick=data,
            enlarge x limits={abs=0.3cm},
            nodes near coords,
            nodes near coords align={vertical},
            ymin=0,ymax=42,
            label style={font=\footnotesize},
            ]
            \addplot[fill=black!80] table[x=interval,y=count]{\mydata};
        \end{axis}
    \end{tikzpicture}
}

\newcommand{\releaseCVC}{
    \pgfplotstableread[row sep=\\,col sep=&]{
        interval & count \\
        0.0.2&4\\
        0.0.3&4\\
        0.0.4&5\\
        0.0.5&5\\
        0.0.6&6\\
        0.0.7&6\\
        0.0.8&5\\
        0.0.11&5\\
        0.0.12&5\\
        1.0.0&5\\
        1.0.1&8\\
        1.0.2&8\\
        trunk&20\\
        }\mydata
    \begin{tikzpicture}
        \tikzset{every node}=[font=\scriptsize\sffamily]
        \begin{axis}[
            ybar,
            bar width=.24cm,
            width=1\textwidth,
            height=4.5cm,
            xticklabel style={rotate=40},
            legend style={at={(0.5,1)},
                anchor=north,legend columns=-1},
            symbolic x coords={0.0.2,0.0.3,0.0.4,0.0.5,0.0.6,0.0.7,0.0.8,0.0.11,0.0.12,1.0.0,1.0.1,1.0.2,trunk},
            xtick pos=left,
            ytick=\empty,
            xtick=data,
            enlarge x limits={abs=0.45cm},
            nodes near coords,
            nodes near coords align={vertical},
            ymin=0,ymax=24,
            label style={font=\footnotesize},
            ]
            \addplot[fill=white] table[x=interval,y=count]{\mydata};
        \end{axis}
    \end{tikzpicture}
}

\newcommand{\comparison}{
    \pgfplotstableread[row sep=\\,col sep=&]{
        interval & count \\
        OpFuzz&3\\
        TypeFuzz&6\\
        STORM&3\\
        YinYang&1\\
        HistFuzz&11\\
        }\mydata
    \begin{tikzpicture}
        \tikzset{every node}=[font=\tiny\sffamily]
        \begin{axis}[
            ybar,
            bar width=.3cm,
            width=\textwidth,
            height=4.5cm,
            xticklabel style={rotate=0},
            legend style={at={(0.5,1)},
                anchor=north,legend columns=-1},
            symbolic x coords={OpFuzz,TypeFuzz,STORM,YinYang,HistFuzz},
            xtick pos=left,
            ytick=\empty,
            xtick=data,
            enlarge x limits={abs=0.5cm},
            nodes near coords,
            nodes near coords align={vertical},
            ymin=0,ymax=14,
            label style={font=\footnotesize},
            ]
            \addplot[fill=white] table[x=interval,y=count]{\mydata};
        \end{axis}
    \end{tikzpicture}
}

\newcommand{\component}{
\pgfplotsset{
tick label style={font=\tiny},
}
\begin{tikzpicture}[scale=1.0]
\begin{axis}[
    ymin=0.2,
    height=3.5cm,
    width=6cm,
    xtick pos=left,
    ytick pos=left,
    enlarge x limits={abs=0.3cm},
    every node near coord/.append style={font=\tiny\sffamily},
    xtick={0,12,24,36,48,60,72,84,96,108,120},
    ytick={2,4,6,8,10,12,14},
    label style={font=\footnotesize},
    ]
\addplot[line width=1pt,black!72,mark options={mark size=1pt},smooth] coordinates {
    (0,0)
    (0.5,2)
    (1,3)
    (1.5,4)
    (3,5)
    (12,6)
    (83.5,7)
    (120,7)
};
\addplot[line width=1pt,red!80,,mark options={mark size=0.8pt},smooth,dash pattern=on 3pt off 2pt on 1pt] coordinates {
    (0,0)
    (0.5,2)
    (1,3)
    (3,4)
    (6,5)
    (26,6)
    (30,7)
    (36.5,8)
    (69.5,9)
    (120,9)
};
\addplot[line width=1pt,blue!80,mark options={mark size=1.5pt},smooth, dashed] coordinates {
    (0,0)
    (0.5,3)
    (1,4)
    (1.5,5)
    (2,6)
    (6,7)
    (17,8)
    (40,9)
    (82,10)
    (120,10)
};
\end{axis}
\end{tikzpicture}
}

\newcommand{\reproduce}{
\pgfplotsset{
tick label style={font=\tiny},
}
\begin{tikzpicture}[scale=1.0]
\begin{axis}[
    ymin=0.2,
    height=4cm,
    width=7cm,
    xtick pos=left,
    ytick pos=left,
    enlarge x limits={abs=0.4cm},
    every node near coord/.append style={font=\tiny},
    xtick={1,2,3,4,5,6,7,8,9,10},
    ytick={12,24,36,48,60,72,84,96},
    ]
\addplot[line width=0.8pt,orange,mark options={mark size=0.9pt},mark=*] coordinates {
    (1,0.4)
    (2,0.54)
    (3,96)
    (4,4.1)
    (5,11.38)
    (6,21.3)
    (7,96)
    (8,4.68)
    (9,96)
    (10,96)
};
\addplot[line width=0.8pt,cyan!70!black,mark options={mark size=0.9pt},mark=*,dashed] coordinates {
    (1,0.2)
    (2,0.46)
    (3,48.95)
    (4,6.8)
    (5,8)
    (6,9.1)
    (7,16.8)
    (8,0.7)
    (9,82.1)
    (10,95)
};

\end{axis}
\end{tikzpicture}
}

\newcommand{\typetable}{
\begin{figure}[t]
    \renewcommand{\arraystretch}{1.2}
    \setlength{\tabcolsep}{3.5pt}
    \centering
    \footnotesize
    \begin{tabular}{ll}   
        \toprule    
        \textbf{Conversion Function Symbol}  &\textbf{Description} \\    
        \midrule          
        \texttt{ite}  & Convert \texttt{Bool} to any sort \\  
        \texttt{is_int} & Convert \texttt{Int} or \texttt{Real} to \texttt{Bool} \\ 
        \texttt{to_real} & Convert \texttt{Int} to \texttt{Real} \\       
        \texttt{str.from_int,str.from_code} & Convert \texttt{Int} to \texttt{Str} \\   
        \texttt{int2bv} & Convert \texttt{Int} to \texttt{BitVec}  \\   
        \texttt{to_int} & Convert \texttt{Real} to \texttt{Int} \\
        \texttt{str.is_digit} & Convert \texttt{Str} to \texttt{Bool} \\    
        \texttt{str.to_int,str.to_code} & Convert \texttt{Str} to \texttt{Int} \\
        \texttt{bv2nat} & Convert \texttt{BitVec} to \texttt{Int} \\

        \bottomrule     
    \end{tabular}
\label{typetable}
\caption{Conversion function symbols defined in SMT-LIB. 
    \label{fig:op-type-table}} 
\end{figure}
}

\newcommand{\bugcounttable}{
{\small
\setlength{\tabcolsep}{0.9em}
\renewcommand{\arraystretch}{1.3}
\begin{tabular}{lrr|r}
\toprule
\textbf{Status} &  \textbf{Z3} &  \textbf{cvc5}  &  \textbf{Total} \\
\midrule
Reported  &  15 &  12  &    27 \\
Confirmed &  10 &   12  &   22 \\
Fixed     &  8 &   8  &    16 \\
Duplicate &  1 &    0  &    1  \\
\bottomrule
\end{tabular}
}
}

\newcommand{\bugtypetable}{
{\small
\setlength{\tabcolsep}{0.8em}
\renewcommand{\arraystretch}{1.3}
\begin{tabular}{lrr|r}
\toprule
\textbf{Type} &  \textbf{Z3} &  \textbf{cvc5}  &  \textbf{Total} \\
\midrule
Correctness &  6 &  2  &    8 \\
Completeness &  3 &   2 &   5 \\
Performance &  1 &   1  &    2 \\
Crash         &  5 &     7  &  12 \\ 
\bottomrule
\end{tabular}
}
}

\newcommand{\bugcomponent}{
{\small
\setlength{\tabcolsep}{1em}
\renewcommand{\arraystretch}{1.3}
\begin{tabular}{lrrr}
\toprule
\textbf{Component} &  \textbf{Z3} &  \textbf{cvc5}  &  \textbf{Total} \\
\midrule
Rewriters &  4 &  5  &    9 \\
Others    &  1 &  3  &    4 \\
Unknown   &  3 &  0  &    3 \\
\midrule
\textbf{All} &  \textbf{8} &  \textbf{8} &  \textbf{16} \\ 
\bottomrule
\end{tabular}
}
}

\newcommand{\newbugstat}{
{
\footnotesize
\setlength{\tabcolsep}{1em}
\renewcommand{\arraystretch}{1.4}
\begin{tabular}{cllll}
\toprule
\textbf{No.} &  \textbf{Solver} &  \textbf{\#ID} &  \textbf{Type} & \textbf{Status} \\
\midrule
1 &  Z3 & 7027 &  crash &  fixed  \\
2 &  Z3 & 7094 & performance & confirmed \\
3 &  Z3 & 7113 & completeness &  reported \\
4 &   Z3 & 7114 & soundness & fixed \\
5 &  cvc5 &  10346 & crash & confirmed \\
6 &  cvc5 &  10373 & completeness & confirmed \\
7 &  cvc5 &  10429 & performance & confirmed \\
8 &  OpenSMT2 & 690 & crash & fixed \\
9 &  OpenSMT2 & 694 & crash & fixed \\
\bottomrule
\end{tabular}
}
}





\newcommand{\validproportion}{
\definecolor{Color1}{RGB}{200,36,35}
\definecolor{Color2}{RGB}{40,120,181}
\begin{tikzpicture}
\pgfplotsset{
tick label style={font=\footnotesize},
}
\begin{axis}[
height=5cm,
width=8cm,
ymin=0.1,
ymax=0.9,
xmin=0.05,
xmax=1.05,
xtick={0.1,0.2,0.3,0.4,0.5,0.6,0.7,0.8,0.9,1.0},
ytick={0.1,0.2,0.3,0.4,0.5,0.6,0.7,0.8},
xlabel={\small \textbf{Temperature}},
ylabel={\small \textbf{Proportion}},
legend style={at={(0.97,0.97)}, anchor=north east},
grid=major,
grid style={dashed,gray!30},
every axis plot/.append style={thick},
]
\addplot[
color=Color2,
mark=*,
mark size=1.5pt,
]
coordinates {
(0.1,0.61)
(0.2,0.57)
(0.3,0.49)
(0.4,0.51)
(0.5,0.55)
(0.6,0.48)
(0.7,0.54)
(0.8,0.51)
(0.9,0.48)
(1.0,0.38)
};

\addplot[
color=Color1, 
mark=square*, 
mark size=1.5pt,
]
coordinates {
(0.1,0.69) 
(0.2,0.64)
(0.3,0.70)
(0.4,0.69)
(0.5,0.63)
(0.6,0.63)
(0.7,0.67)
(0.8,0.67)
(0.9,0.73)
(1.0,0.68)
};

\end{axis}
\end{tikzpicture}
}

\newcommand{\zthreetrace}{
\pgfplotsset{
tick label style={font=\tiny},
}
\begin{tikzpicture}[scale=1.0]
\begin{axis}[
    title={\scriptsize Z3},
    ymin=0.03, 
    ymax=0.11, 
    height=4cm,
    width=9cm,
    xtick pos=left,
    ytick pos=left,
    enlarge x limits={abs=0.3cm},
    symbolic x coords={1, 2, 3, 4, 5, 6, 7, 8, 9, 10, 11, 12, 13, 14, 15, 16, 17, 18, 19, 20, 21, 22, 23, 24, 25, 26, 27, 28, 29, 30},
    every node near coord/.append style={font=\tiny},
    xtick={1,5,10,15,20,25,30},
    ytick={0.04,0.06,0.08,0.10},
    yticklabels={0.04,0.06,0.08,0.10},
    ]
\addplot[black!80, mark=x] coordinates {
    (1,0.10084984036712032) (2,0.08506987566297661) (3,0.08066693268643364) (4,0.07795014447903878)
    (5,0.07525469482857287) (6,0.07249928818544482) (7,0.06970200703287002) (8,0.0675746346715247)
    (9,0.06543048839651928) (10,0.0629111761219678) (11,0.060924014368785126) (12,0.058220392988107834)
    (13,0.0565240438088326) (14,0.054324281239315236) (15,0.05366857411941624) (16,0.05063957806099672)
    (17,0.04814282938393053) (18,0.046994776934960444) (19,0.04493111249106707) (20,0.04212960238371542)
    (21,0.041365896136761296) (22,0.038971109568034755) (23,0.037584021587930234) (24,0.036736940367952046)
    (25,0.036385556869548794) (26,0.03535739936610979) (27,0.03575246179743023) (28,0.036565525238789656)
    (29,0.035710102271971925) (30,0.03643769125461813) 
};
\end{axis}
\end{tikzpicture}
}

\newcommand{\cvcfourtrace}{
\pgfplotsset{
tick label style={font=\tiny},
}
\begin{tikzpicture}[scale=1.0]
\begin{axis}[
    ymin=0, 
    ymax=6, 
    height=4cm,
    width=9cm,
    xtick pos=left,
    ytick pos=left,
    enlarge x limits={abs=0.3cm, upper},
    symbolic x coords={0, 1, 2, 3, 4, 5, 6, 7, 8, 9, 10, 11},
    every node near coord/.append style={font=\tiny},
    xtick={0,1,2,3,4,5,6,7,8,9,10,11},
    ytick={0,1,2,3,4,5,6},
    yticklabels={0,1,2,3,4,5,6},
    ]
\addplot[black!80, mark=x] coordinates { (0,0)
(1,0.5301337258731064) (2,1.0246644212819982) (3,1.6146680547207697) (4,2.055057565986598) 
(5,2.626149953961268) (6,3.2910948249707768) (7,3.6378711534984207) (8,4.331514207830253) 
(9,4.3928391995373195) (10,5.251427109407339) 
};
\end{axis}
\end{tikzpicture}
}

\newcommand{\cvcablation}{
\pgfplotsset{
tick label style={font=\tiny},
}
\begin{tikzpicture}[scale=1.0]
\begin{axis}[
    ymin=0, 
    ymax=6, 
    height=4cm,
    width=9cm,
    xtick pos=left,
    ytick pos=left,
    enlarge x limits={abs=0.3cm, upper},
    symbolic x coords={0, 1, 2, 3, 4, 5, 6, 7, 8, 9, 10, 11},
    every node near coord/.append style={font=\tiny},
    xtick={0,1,2,3,4,5,6,7,8,9,10,11},
    ytick={0,1,2,3,4,5,6},
    yticklabels={0,1,2,3,4,5,6},
    ]
\addplot[black!80, mark=x] coordinates { (0,0)
(1,0.5301337258731064) (2,1.0246644212819982) (3,1.6146680547207697) (4,2.055057565986598) 
(5,2.626149953961268) (6,3.2910948249707768) (7,3.6378711534984207) (8,4.331514207830253) 
(9,4.3928391995373195) (10,5.251427109407339) 
};
\addplot[blue, mark=square*] coordinates { (0,0)
(1,1.7139484274612231) (2,1.2201306690345675) (3,1.6139860645514787) (4,1.4536748798370505) 
(5,2.135266376278116) (6,2.712693135964942) (7,3.0007067816637276) (8,3.6040756809908827) 
(9,4.3799925702992) (10,5.169714502407226) 
};
\end{axis}
\end{tikzpicture}
}

\newcommand{\zthreetraceablation}{
\pgfplotsset{
tick label style={font=\tiny},
}
\begin{tikzpicture}[scale=1.0]
\begin{axis}[
    title={\scriptsize Z3},
    ymin=0.02, 
    ymax=0.11, 
    height=4cm,
    width=13cm,
    xtick pos=left,
    ytick pos=left,
    enlarge x limits={abs=0.3cm},
    symbolic x coords={1, 2, 3, 4, 5, 6, 7, 8, 9, 10, 11, 12, 13, 14, 15, 16, 17, 18, 19, 20, 21, 22, 23, 24, 25, 26, 27, 28, 29, 30},
    every node near coord/.append style={font=\tiny},
    xtick={1,5,10,15,20,25,30},
    ytick={0.04,0.06,0.08,0.10},
    yticklabels={0.04,0.06,0.08,0.10},
    ]
\addplot[black!80, mark=x] coordinates {
    (1,0.08351139998615056) (2,0.08025670291212984) (3,0.07814510300297768) (4,0.07486500420092455)
    (5,0.07163674367383792) (6,0.06937455715881476) (7,0.06827154190772881) (8,0.06527556922759008)
    (9,0.06263446949219713) (10,0.0600336159884263) (11,0.057570026846158) (12,0.055916640401465385)
    (13,0.05302415101766513) (14,0.05150609797374882) (15,0.04886985430330796) (16,0.04657527632081631)
    (17,0.045166334316845934) (18,0.04236946926858727) (19,0.03904696993941542) (20,0.03738406484200169)
    (21,0.03500349974669293) (22,0.03193935929653455) (23,0.029855079540255375) (24,0.02799854505691987)
    (25,0.027524578474172067) (26,0.02660306403380814) (27,0.02532187653691671) (28,0.025083070259365573)
    (29,0.0246960707178589) (30,0.025740209865768467)
};
\end{axis}
\end{tikzpicture}
}

\newcommand{\cvcfourtraceablation}{
\pgfplotsset{
tick label style={font=\tiny},
}
\begin{tikzpicture}[scale=1.0]
\begin{axis}[
    title={\scriptsize cvc5},
    ymin=0.10, 
    ymax=0.32, 
    height=4.5cm,
    width=13cm,
    xtick pos=left,
    ytick pos=left,
    enlarge x limits={abs=0.3cm},
    symbolic x coords={1, 2, 3, 4, 5, 6, 7, 8, 9, 10, 11, 12, 13, 14, 15, 16, 17, 18, 19, 20, 21, 22, 23, 24, 25, 26, 27, 28, 29, 30},
    every node near coord/.append style={font=\tiny},
    xtick={1,5,10,15,20,25,30}, 
    ytick={0.10,0.15,0.20,0.25,0.30,0.35}, 
    yticklabels={0.10,0.15,0.20,0.25,0.30,0.35}, 
    ]
\addplot[black!80, mark=x] coordinates {
(1,0.28824691029411653) (2,0.2798624934334537) (3,0.2694057483087204) (4,0.26040054484992176)
(5,0.2522972235718606) (6,0.2472557128368702) (7,0.23899759155824835) (8,0.23233284100956966)
(9,0.22278644664020308) (10,0.21866326250644424) (11,0.21604393643551859) (12,0.20423856325835818)
(13,0.19919671978396725) (14,0.19225591798572253) (15,0.1875219153115572) (16,0.17695931270176704)
(17,0.179052543749799) (18,0.17265773760753136) (19,0.16789096458605957) (20,0.1621227498533362)
(21,0.1583463209997257) (22,0.15524425483773874) (23,0.1623375176292146) (24,0.16110443293938168)
(25,0.1554250672877269) (26,0.16085788668749823) (27,0.15032887536204223) (28,0.15661155743040775) (29,0.1536813854866602) (30,0.1524907166788276) 
};
\end{axis}
\end{tikzpicture}
}

%% file: content/introduction.tex
As software systems grow in scale and complexity, critical security vulnerabilities have become increasingly prevalent~\cite{blogReviewZeroday, cve}. 
For instance, in crucial system software like the Linux kernel~\cite{DBLP:conf/apsys/ChenMWZZK11}, over 200 security vulnerabilities are disclosed each year on average~\cite{DBLP:journals/tosem/JiangJWMLZ24}.
This challenge is further compounded by the rapid adoption of AI-assisted code generation, which frequently introduces subtle flaws into production codebases~\cite{DBLP:journals/corr/abs-2512-03262}. 
While automated vulnerability detection techniques, most notably fuzz testing, have successfully scaled to identify thousands of bugs in production systems~\cite{serebryany2017oss, manes2019fuzzingSurvey}, the subsequent mitigation phase remains a severe bottleneck. 
Resolving deep-rooted software crashes, particularly those involving memory corruption or complex state violations, fundamentally relies on domain expertise and exhaustive manual debugging. 
Consequently, the widening gap between vulnerability discovery and patch deployment imposes substantial security risks and maintenance costs~\cite{anwar2020measuring, iannone2022secret, cybersecurityventuresCybercrimeCost}.

To address this mitigation bottleneck, the research community has increasingly explored Automated Program Repair (APR)~\cite{liu2019tbar, le2012GenProg} and LLM-based software engineering agents~\cite{xia2024chatRepair, yin2024thinkrepair, bouzenia2024repairagent, zhang2025ReinFix}. 
By leveraging the code-comprehension capabilities of LLMs, these frameworks attempt to autonomously localize faults and synthesize patches. 
However, despite achieving impressive results on standard benchmarks~\cite{jimenez2023swebench}, current systems fundamentally treat vulnerability repair as a static text-generation problem~\cite{DBLP:conf/nips/YangJWLYNP24, wang2024openhands, xia2025agentless}. 
They typically rely on static artifacts, such as issue reports, basic stack traces, and source code, without reasoning about the program's dynamic execution state. While this static approach is effective for relatively straightforward bugs in high-level languages (e.g., Python projects in SWE-bench~\cite{DBLP:conf/iclr/JimenezYWYPPN24}), it struggles to resolve complex, real-world security vulnerabilities that demand a nuanced comprehension of dynamic behavior and memory layouts~\cite{lee2025secbench}.

In practice, this reliance on static analysis stands in stark contrast to how systems experts debug complex vulnerabilities. 
When human developers encounter a native crash, they do not merely analyze a static codebase.
Instead, they employ interactive debuggers (e.g., \texttt{GDB}) to inspect raw memory and register states,
augmentation tools (e.g., \texttt{pwndbg}~\cite{pwndbg}) for memory and heap introspection, and deterministic replay (e.g., \texttt{rr}~\cite{DBLP:conf/usenix/OCallahanJFHNP17}) to trace the propagation of corrupted states from the crash site back to the root cause. 
Moreover, distinct vulnerability classes, such as heap corruptions or use-after-free errors, demand potentially different debugging strategies and tool invocations. 
Even recent efforts~\cite{yu2025patchagent,DBLP:journals/corr/abs-2601-13933} that attempt to mimic human debugging workflows with LLMs still largely relegate the agent to the role of a passive code reader, limited to language server queries that fetch static code information without any dynamic execution context.
By overlooking this interactive, dynamic debugging process, existing LLM-based agents may miss critical execution context. 
We argue that the rich, dynamic state derived from live execution is essential to resolve non-trivial memory safety crashes.

\mypara{Our Approach.}
To address this gap, we introduce \tool, an autonomous debugging harness\footnote{In the context of LLM agents, a \textit{harness} provides an operating system-like abstraction~\cite{schmid2026agent}. It treats the language model as the central reasoning processor while managing the execution state (analogous to memory) and exposing standardized interfaces (drivers) for external tool invocation.} that orchestrates LLM-driven analysis through dynamic execution state introspection.
\tool\ operates as an end-to-end harness built on two core mechanisms:  \textit{signature-driven investigation} and \textit{interactive state introspection}.
Specifically, the workflow initiates with a reproducible Proof-of-Concept (PoC) trigger, typically sourced from a fuzzing campaign or an issue report. 
By analyzing the initial crash signature, \tool\ classifies the vulnerability type and dynamically injects type-specific debugging guidelines into the LLM's context. 
This \textit{signature-driven investigation} strategy directs the agent's attention to the most relevant memory states and execution traces, effectively mirroring the specialized approaches of human experts.

During the diagnostic phase, \tool\ is allowed to map sanitizer trap information to the corresponding source code via a language server, which provides the necessary static context for the LLM.
Additionally, to establish a dynamic understanding of the program state, the \textit{interactive state introspection} component empowers the LLM to orchestrate a specialized suite of systems tools. 
This includes standard debuggers (e.g., \texttt{GDB}) for execution control, debugger augmentation tools (e.g., \texttt{pwndbg}) for deep memory and heap introspection, and record-and-replay systems (e.g., \texttt{rr}) for deterministic reverse execution. 
Guided by the signature-driven investigation strategy, the agent iteratively formulates hypotheses regarding the root cause and actively interacts with these tools to verify its assumptions if needed. 
By bridging the gap between static information and dynamic introspection, \tool\ acquires the precise contextual information required to diagnose memory safety issues.

Ultimately, upon pinpointing the root cause, \tool\ synthesizes a patch and attempts validation by recompiling the target application and re-executing the PoC trigger. 
Should the patch fail validation, \tool\ seamlessly feeds the resulting compiler warnings or subsequent crash logs back into the agent's context, initiating a new refinement iteration. This structured, closed-loop workflow grounds the LLM's reasoning in empirical, dynamic observations. 
Consequently, this structured, closed-loop workflow mitigates divergent agent behavior (e.g., hallucination) and enables \tool\ to successfully resolve intricate vulnerabilities that defeat state-of-the-art static LLM agents.

\mypara{Results.}
To evaluate the efficacy of \tool, we conducted an extensive empirical study using SEC-bench~\cite{lee2025secbench}, a rigorous benchmark comprising 200 real-world security vulnerabilities across 29 popular open-source C/C++ projects. 
Our evaluation demonstrates that by reasoning over dynamic memory states and execution traces, \tool\ successfully resolves complex vulnerabilities that evade state-of-the-art static LLM agents. 
Specifically, \tool\ achieves a resolution rate of approximately \textbf{90\%} across the benchmark, outperforming well-known baselines, \eg\ PatchAgent~\cite{yu2025patchagent} and VulnResolver~\cite{DBLP:journals/corr/abs-2601-13933}, which achieve resolution rates of 57.5\% and 67.5\%, respectively.
Furthermore, our ablation studies confirm that both core components, the signature-driven investigation and interactive state introspection, are critical to achieving this high resolution rate. 
These findings validate our core hypothesis: dynamic, interactive debugging is strictly necessary for autonomously resolving real-world, memory-safety vulnerabilities.

In summary, \tool\ represents a fundamental shift in automated program repair, bridging the gap between LLM-based code reasoning and the dynamic complexity of systems-level software. The main contributions are:

\begin{itemize}[leftmargin=*,label=$\star$]
    \item \textbf{Interactive LLM Debugging Harness:} We propose a transition from static codebase analysis to dynamic, interactive execution debugging for automated vulnerability repair. \tool\ innovatively integrates interactive debugging actions directly into the LLM's reasoning loop, providing the rich execution context necessary to diagnose complex system failures.
    
    \item \textbf{\tool\ System Implementation:} We design and implement an end-to-end framework that orchestrates an LLM through a structured, closed-loop debugging process. By guiding exploration via signature-driven investigation and interactive state introspection, \tool\ effectively isolates root causes while grounding the model's reasoning to mitigate non-deterministic behavior.
    
    \item \textbf{Comprehensive Evaluation:} We evaluate \tool\ against a rigorous benchmark of real-world C/C++ vulnerabilities. Our results demonstrate that reasoning over dynamic program states enables \tool\ to successfully localize and repair critical bugs that defeat state-of-the-art baselines, significantly advancing the capabilities of automated vulnerability resolution.

\end{itemize}

\mypara{Organization.}
The rest of the paper is organized as follows.
Section~\ref{sec:motivation} provides background context and a motivating example that illustrates the limitations of static LLM agents and the necessity of dynamic debugging. 
Section~\ref{sec:design} details the design and implementation of \tool, including its core components and debugging workflow. 
Section~\ref{sec:evaluation} describes our experimental setup and presents the results of our evaluation on the SEC-bench dataset. 
Section~\ref{sec:discussion} discusses the implications of our findings, limitations of \tool, and potential avenues for future work.
Finally, Section~\ref{sec:related-work} reviews related research in automated program repair and LLM-based software engineering agents, and Section~\ref{sec:conclusion} concludes the paper with a summary of our contributions.

%% file: content/motivation.tex
In this section, we first provide background on the automated program repair (APR) landscape and the limitations of existing LLM-based approaches. We then present a motivating example that illustrates the critical need for dynamic execution context in resolving complex vulnerabilities, motivating the design of \tool.

\begin{figure*}[t]

\begin{subfigure}{1\linewidth}
\includegraphics[width=\linewidth]{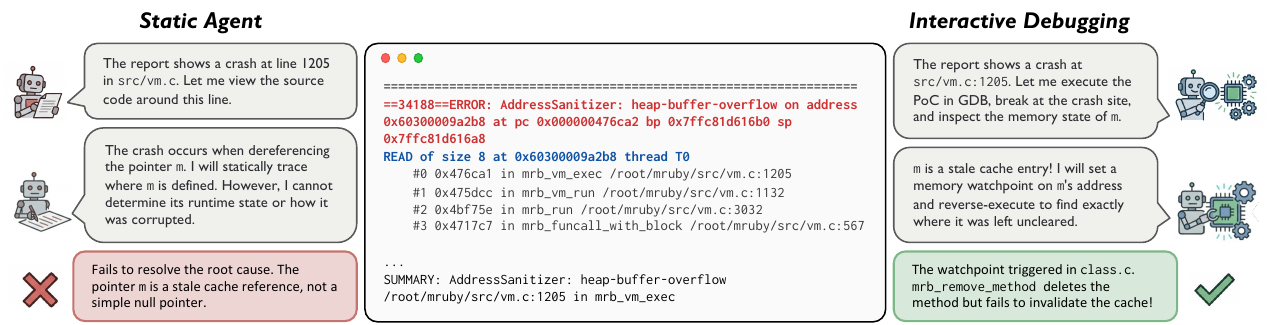}
\vspace{0.5em}
\caption{ASan report of CVE-2022-1286 showing the crash site and the workflows of a static agent and interactive debugging agent.}
\label{fig:cve2022_1286_asan}
\end{subfigure}

\vspace{1em}

\begin{subfigure}[t]{0.43\linewidth}
\begin{lstlisting}[language=C,frame=none,morekeywords={MRB_METHOD_NOARG_P,MRB_METHOD_FUNC}]
// --- [Symptom]: src/vm.c ---
// Indirect call executing stale method
/* ... */
if (MRB_METHOD_PROC_P(m)) {
  /* ... */
}
else {
  if (MRB_METHOD_NOARG_P(m)) {
    check_method_noarg(mrb, ci);
  }
  recv = <@\textbf{MRB\_METHOD\_FUNC(m)(mrb, recv)}@>; 
}
\end{lstlisting}
\vspace{-0.5em}
\caption{Symptom: Indirect call in \texttt{vm.c}}
\label{lst:cve2022_1286_symptom}
\end{subfigure}
\begin{subfigure}[t]{0.55\linewidth}
\begin{lstlisting}[language=C,frame=none,morekeywords={mrb_malloc}]
// --- [Root Cause]: src/class.c ---
void mrb_remove_method(mrb_state *mrb, 
                     struct RClass *c, mrb_sym mid) {
  /* ... */
  // Vulnerable logic: Cache is not cleared 
  // after method removal
  if (h && mt_del(mrb, h, mid)) {
+  <@\colorbox{orange!30}{\textbf{ mrb\_mc\_clear\_by\_class(mrb, c);}}@>
    return;
  }
  /* ... */
}
\end{lstlisting}
\vspace{-0.5em}
\caption{Root cause: Missing cache invalidation in \texttt{class.c}}
\label{lst:cve2022_1286_rootcause}
\end{subfigure}

\caption{CVE-2022-1286 code snippets showing (a) a comparison of resolution workflows. Static agents are limited to analyzing the ASan report and source code, leading them to stall at the symptom site and propose superficial patches. In contrast, \tool\ mimics true human debugging by utilizing dynamic memory introspection and watchpoints to trace the stale pointer back to its root cause in a different file. (b) The symptomatic indirect call in \texttt{vm.c}. (c) The actual root cause and patch involving missing cache invalidation in \texttt{class.c}.}
\label{fig:cve2022_1286}
\end{figure*}

\subsection{Background}
\label{sec:background}

Automated Program Repair (APR) seeks to reduce the manual effort required to diagnose and resolve software defects~\cite{zhang2023learningBasedAPRSurvey}. In this work, we target a \textit{PoC-driven} repair setting: the system is provided with a proof-of-concept (PoC) input that triggers the vulnerability, a crash description (\eg a sanitizer report), and a functional test suite that guards the program's core logic. This setting naturally aligns with modern continuous fuzzing infrastructures, such as OSS-Fuzz~\cite{serebryany2017oss}, which routinely produce PoC inputs alongside crash reports. Addressing this scenario is critical, as the gap between automated bug \textit{discovery} and automated bug \textit{resolution} continues to widen~\cite{yu2025patchagent}.

The traditional PoC-driven repair pipeline consists of three stages: fault localization, patch generation, and patch validation. Classical fault localization techniques, ranging from symbolic execution~\cite{10.1145/3052973.3053033, DBLP:conf/uss/He0S0L22} to spectrum-based scoring~\cite{DBLP:conf/uss/BlazytkoSAAFWH20}, often incur substantial computational overhead or yield imprecise candidate lists. Similarly, traditional patch generators (whether search-, constraint-, or pattern-based~\cite{DBLP:conf/wcre/LeLG16, DBLP:journals/tosem/GaoWDJXR21, DBLP:conf/icse/LongR16}) trade expressiveness against the size of the search space. Holistic solutions, such as ExtractFix~\cite{DBLP:journals/tosem/GaoWDJXR21}, attempt to chain these stages end-to-end using symbolic execution. However, they suffer from severe path-explosion overhead and struggle to synthesize fixes for complex memory-safety violations, such as use-after-free (UAF) errors.

Given their strong capabilities in code comprehension and generation~\cite{DBLP:conf/nips/BrownMRSKDNSSAA20}, Large Language Models (LLMs) have emerged as a natural evolution for APR~\cite{DBLP:conf/sp/PearceTAKD23a}. 
Nonetheless, the vast majority of existing LLM-based tools address only a single stage of the pipeline. 
For example, some systems utilize multi-turn LLM interactions for fault localization~\cite{DBLP:journals/tse/TuZJYLJ24}, possibly with human-in-the-loop feedback~\cite{DBLP:conf/gecco/MurtazaMRMB24}, while others focus purely on iteratively improving candidate patches given prior validation feedback~\cite{DBLP:conf/aiware/KulsumZXd24}. 
Even systems that integrate generation and validation often assume the exact fault location is already known, avoiding the most challenging aspect of the repair process entirely~\cite{DBLP:conf/sp/PearceTAKD23a}.
Subsequently, researchers have attempted to unify the entire pipeline under a single LLM workflow. 
For instance, Agentless~\cite{xia2025agentless} employs a single LLM to perform both fault localization and patch generation and achieves impressive results on the SWE-bench~\cite{DBLP:conf/nips/YangJWLYNP24} benchmark of Python issues.
Moreover, to address the inherent challenges of debugging complex vulnerabilities, PatchAgent~\cite{yu2025patchagent} is devised to mimic human debugging workflows by allowing the LLM to query a language server for static code navigation.
However, PatchAgent's design still treats vulnerability repair as a purely static problem. 
The agent is limited to analyzing static artifacts: the crash report, stack trace, and source code, without any access to dynamic execution context. This contrasts with how human experts approach debugging in practice and misses critical information necessary to resolve complex native crashes.

\subsection{Motivating Example}
\label{sec:motivating_example}

To illustrate the necessity of dynamic runtime introspection for APR, we examine CVE-2022-1286~\cite{cve-2022-1286}, a complex heap buffer overflow in the \texttt{mruby} interpreter, as depicted in Figure~\ref{fig:cve2022_1286}.
Existing vulnerability resolution techniques typically rely on static source code analysis augmented by sanitizer crash reports. 
When applied to this vulnerability, static LLM agents successfully parse the ASan trace to identify the crash site in \texttt{vm.c}. 
However, their reasoning stalls at the faulting indirect method invocation (Figure~\ref{lst:cve2022_1286_symptom}).
Because the actual root cause, a failure to invalidate the method cache during a prior method removal operation in \texttt{class.c} (Figure~\ref{lst:cve2022_1286_rootcause}), occurs much earlier in the execution trace, the responsible function (\texttt{mrb\_remove\_method}) is entirely absent from the ASan stack trace. 

Restricted to this static context, LLMs lack the temporal visibility required to connect the memory corruption back to the stale pointer's origin. 
Recent prior work, such as PatchAgent~\cite{yu2025patchagent}, attempts to overcome this limitation by mimicking human experts through static code navigation (e.g., following variable definitions across files). 
Yet, navigating definitions statically is insufficient when the control flow involves dynamic caching and indirect state mutations. 
Consequently, as shown in the left workflow of Figure~\ref{fig:cve2022_1286_asan}, this structural limitation restricts its ability to trace the root cause, leading to stalled reasoning that fails to connect the symptom to the underlying issue.
Notably, PatchAgent explicitly documents CVE-2022-1286 as a known failure case, acknowledging that their lack of dynamic execution context prevents the resolution of such complex issues~\cite{yu2025patchagent}.

In practice, human developers do not resolve complex memory corruptions by merely reading static crash traces and cross-referencing function definitions. 
Instead, they typically rely on interactive runtime state tracking. 
A developer executes the PoC trigger within a debugger, sets a watchpoint on the corrupted memory address, and traces the program state backward to discover exactly where and how the stale pointer was introduced. 
By equipping an LLM agent with these exact interactive debugging primitives, \tool\ bridges the spatial and temporal gap between symptom and root cause. 
Thus, as illustrated in the right workflow of Figure~\ref{fig:cve2022_1286_asan}, \tool\ dynamically intercepts the program state immediately prior to the faulting indirect call. 
The agent inspects the runtime values of the target method structure \texttt{m} and determines it is a stale reference fetched silently from the internal method cache. 
This dynamic insight provides conclusive evidence that the vulnerability stems from a stale cache entry, rather than a localized buffer sizing error. 
By utilizing watchpoints to track the lifecycle of the cached reference backward through the execution flow, the agent identifies the precise origin of the stale pointer: the missing \texttt{mrb\_mc\_clear\_by\_class} invalidation call in \texttt{src/class.c}. 
Ultimately, this example demonstrates that emulating the dynamic, state-aware investigative strategies of human experts is indispensable for diagnosing and repairing complex vulnerabilities in real-world systems software.

%% file: content/approach.tex
\begin{figure*}[t]
\centering
\includegraphics[width=0.93\linewidth]{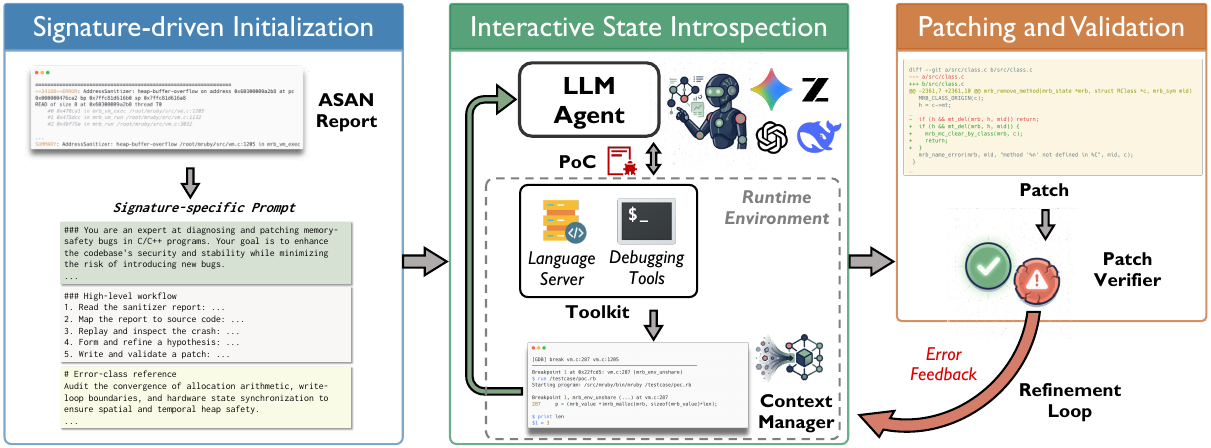}
\caption{Overview of \tool's workflow.}
\label{fig:workflow}
\end{figure*}

In this section, we present the design and implementation of \tool, an autonomous debugging agent harness that leverages LLMs to resolve complex memory-safety vulnerabilities in C/C++ software. 
We first outline the high-level architecture of \tool\ and its core components, then detail the agentic debugging workflow that orchestrates the interaction between the LLM and dynamic system tools, and finally describe our implementation choices and optimizations to enable robust, scalable debugging.

\subsection{Overview}
\label{sec:overview}

\tool\ orchestrates a structured, closed-loop debugging workflow that emulates the investigative heuristics of human experts, as illustrated in Figure~\ref{fig:workflow}. 
Specifically, \tool\ starts with a reproducible crash trigger, such as a PoC input that reliably induces a memory-safety vulnerability reported by a sanitizer, and operates in three distinct phases.

First, \tool\ performs \textit{signature-driven initialization}, where it parses the initial crash report to identify the error class and injects tailored troubleshooting guidelines into the LLM's system prompt (\S\ref{sec:workflow_phase1}). 
Next, it transitions to \textit{interactive state introspection} (\S~\ref{sec:workflow_phase2}). 
Here, \tool\ maps the runtime crash context to static source locations and enters an iterative hypothesis-testing loop. 
The agent commands debuggers to inspect memory and trace execution flow, relying on data summarization and deterministic time-travel debugging to manage context size and safely navigate program state. 
Finally, once the root cause is isolated, \tool\ proceeds to \textit{patching and validation} (\S~\ref{sec:workflow_phase3}), synthesizing a patch and validating it against the PoC in a closed feedback loop. 
After each phase, the system appends any new findings (e.g., debugger outputs, compiler warnings) to the agent's context, allowing it to iteratively refine its understanding and approach until a verified patch is synthesized.

\begin{figure}
\centering
\includegraphics[width=0.9\linewidth]{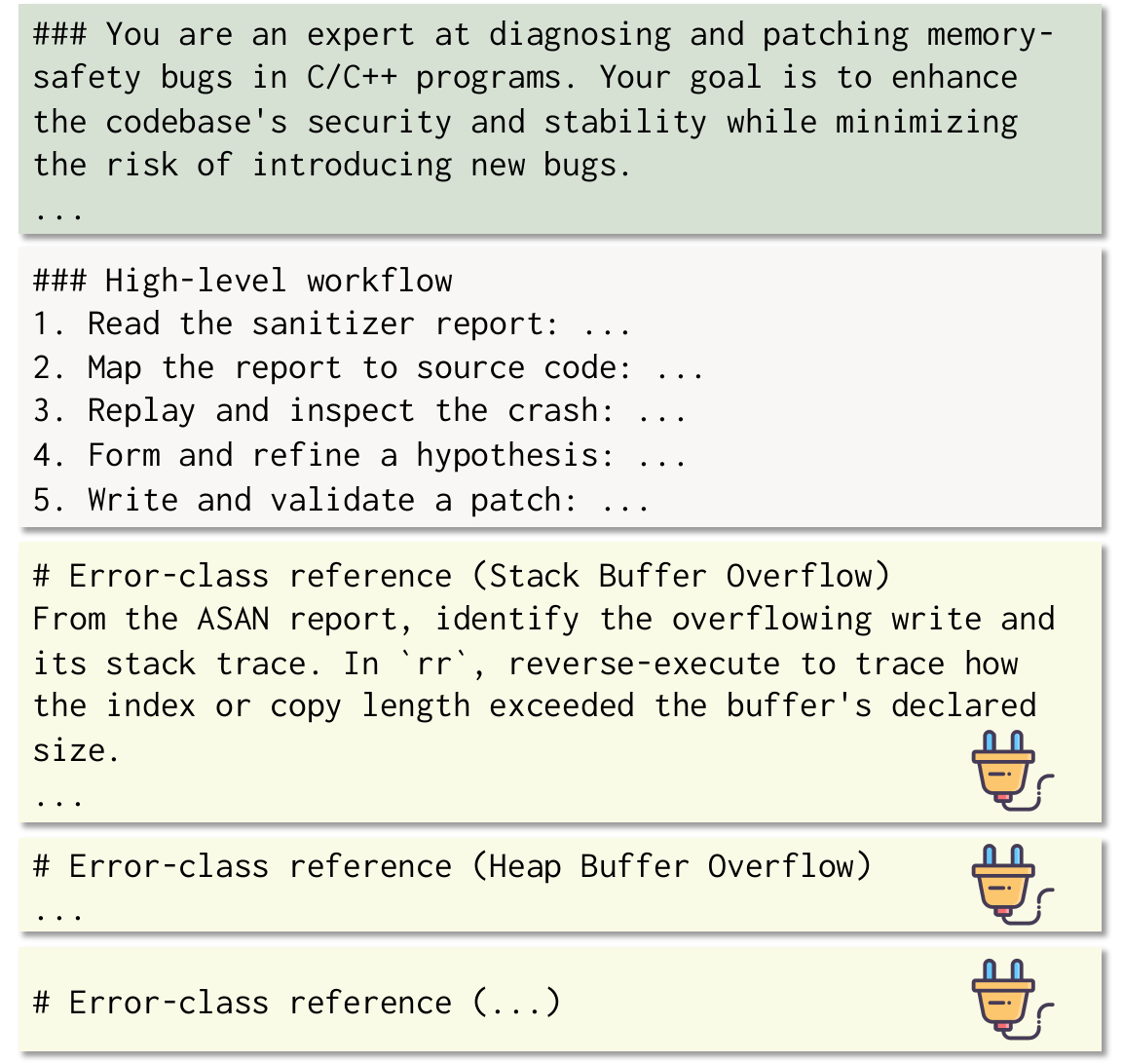}
\caption{The prompt template for signature-driven initialization. \tool\ injects error-class-specific troubleshooting guidelines based on the crash signature.}
\label{fig:prompt}
\end{figure}

\subsection{Signature-Driven Initialization}
\label{sec:workflow_phase1}

Algorithm~\ref{algo:debugagent} outlines the high-level workflow of \tool.
The workflow begins when a runtime sanitizer (e.g., ASan) traps a memory-safety violation triggered by the PoC input.
As depicted in Algorithm~\ref{algo:debugagent}, \tool\ initially takes the buggy project $P$, its test suite $T$, the PoC trigger $I_{poc}$, and the sanitizer report $R_{san}$ as input.
To constrain the LLM's non-deterministic search space and prevent premature patch synthesis, \tool\ employs a signature-driven initialization strategy (Line~\ref{algo:phase1}).
By parsing the sanitizer report, the system extracts the specific vulnerability class (e.g., \texttt{use-after-free} or \texttt{heap-buffer-overflow}) and the trapping instruction. 
Based on this signature, \tool\ then selectively injects domain-specific troubleshooting guidelines into the LLM's system prompt, as illustrated in Figure~\ref{fig:prompt}.
These guidelines incorporate error-specific heuristics derived from expert knowledge bases~\cite{DBLP:books/daglib/0039904}, detailing common root causes and recommended debugging strategies. 
This injection also enforces strict rules of engagement: the LLM must articulate a concrete hypothesis regarding the root cause and is explicitly penalized for proposing code modifications before dynamically verifying its assumptions. 
Furthermore, the guidelines steer the investigation based on the error class; for instance, diagnosing a \texttt{use-after-free} prioritizes heap metadata inspection and execution tracing, whereas a \texttt{heap-buffer-overflow} directs focus toward boundary checks and buffer allocation logic.

\subsection{Interactive State Introspection\label{sec:workflow_phase2}}

Equipped with the initial crash signature and troubleshooting guidelines, the agent must bridge the runtime crash context and the static source logic. 
Specifically, \tool\ launches an interactive debugging session (Line~\ref{algo:init_session}) and enters a closed debugging loop, attempting to identify the root cause and synthesize a patch to fix the bug (Lines~\ref{algo:loop}--\ref{algo:patch:end}).
To enable this, \tool\ first instructs the LLM to analyze the current crash context and formulate a preliminary hypothesis regarding the root cause.
Additionally, it should propose concrete debugging steps to verify its assumptions, such as fetching specific code snippets or querying the execution state with a debugger.
To verify its assumptions, the agent may delegate search queries to a language server, which maps the stack frames and memory symbols from the sanitizer report to concrete source code locations (Lines~\ref{algo:view_source:start}--\ref{algo:view_source:end}).
This mapping exposes the relevant code snippets surrounding the crash site or other suspicious locations, providing the static context necessary to establish initial breakpoints or memory watchpoints.
Moreover, \tool\ is capable of issuing commands to debuggers to inspect memory state and trace execution flow as needed (Lines~\ref{algo:llm_query}--\ref{algo:execute_debugger:end}).
Notably, standard debuggers (e.g., \texttt{GDB}) are fundamentally forward-only; an erroneous step by the agent can irreversibly mutate program state, forcing a costly session restart. 
To complement the standard debugging capabilities, \tool\ integrates Mozilla \texttt{rr}~\cite{DBLP:conf/usenix/OCallahanJFHNP17} for deterministic record-and-replay debugging, alongside \texttt{pwndbg}~\cite{pwndbg} for deep memory introspection. 
In particular, record-and-replay capabilities allow the agent to issue reversible commands (e.g., \texttt{reverse-continue}) to precisely locate the chronological origin of memory corruptions without destroying the crash context. 
This is particularly crucial for diagnosing temporal memory bugs, such as double-frees.
To prevent invalid debugging commands resulting from LLM hallucination, \tool\ incorporates concrete debugging interfaces and command validation layers.
In this way, the agent can iteratively refine its hypothesis based on empirical evidence from the execution state and relevant source code.

\normalem
\begin{algorithm}[t]
\small
\caption{\small \tool's pseudocode}
\label{algo:debugagent}
\DontPrintSemicolon

\SetKwProg{Fn}{Function}{:}{}
\SetKwFunction{DebugAgent}{DebugHarness}

\SetKwFunction{ExtractGuidelines}{ExtractGuidelines}
\SetKwFunction{MapToSource}{MapToSource}
\SetKwFunction{InitDebugger}{InitDebugger}
\SetKwFunction{LLMQuery}{LLMQuery}
\SetKwFunction{ExecuteDebugger}{ExecuteDebugger}
\SetKwFunction{SummarizeState}{Distill}
\SetKwFunction{SynthesizePatch}{GeneratePatch}
\SetKwFunction{ApplyPatch}{ApplyPatch}
\SetKwFunction{Validate}{Validate}

\SetKwFunction{HasBudget}{HasBudget}
\SetKwFunction{ViewSource}{ViewSource}

\Fn{\DebugAgent{$P$, $T$, $I_{poc}$, $R_{san}$}}{
    \textit{\color{purple} \# Stage 1: Signature-Driven Initialization\;}
    $ctx \leftarrow \ExtractGuidelines(R_{san})$ \label{algo:phase1} \;

    \textit{\color{purple} \# Stage 2: Interactive State Introspection\;}
    $session \leftarrow \InitDebugger(P, I_{poc})$ \label{algo:init_session} \;

    \textit{\color{purple} \# Closed-Loop Debugging\;}
    \While{\textsc{Not}(BugResolved) \textsc{and} \HasBudget() \label{algo:loop}}{
        $cmds, scripts \leftarrow \LLMQuery(ctx)$ \label{algo:llm_query} \;
        \If{$cmds$ \textsc{is} $\text{ViewSource}$ \label{algo:view_source:start}}{
            $ctx \leftarrow ctx \cup \ViewSource(cmds)$ \label{algo:view_source:end}\;
        }
        \If{$cmds$ \textsc{is} $\text{ExecuteDebugger}$ \label{algo:execute_debugger:start}}{
            $out_{raw} \leftarrow \ExecuteDebugger(session, cmds)$ \label{algo:execute_debugger:end}\;
        }
        \textit{\color{purple} \# Local execution of LLM-emitted scripts if any\;}
        $ctx\leftarrow ctx \cup \SummarizeState(ctx, out_{raw}, scripts)$ \label{algo:summarize_state} \;

    \textit{\color{purple} \# Stage 3:  Patching and Validation\;}

    \If{RootCauseFound \label{algo:patch:start}}{
    
        $P' \leftarrow \ApplyPatch(P, \SynthesizePatch(ctx))$ \label{algo:apply_patch}\;
        $status, feedback \leftarrow \Validate(P', I_{poc}, T)$ \label{algo:validate}\;
        
        \If{$status == \textsc{Pass}$}{
            \Return $P'$ \label{algo:return_patch}\;
        }

        \Else {        
        \textit{\color{purple} \# Append failures to refine the next iteration\;}
        $ctx \leftarrow ctx \cup \SummarizeState(ctx, feedback)$ \label{algo:patch:end}\;
        }
    }
    }
    
    \Return $\text{Null}$ \label{algo:return_null}\;
}
\end{algorithm}

A key challenge during this phase is context management.
In particular, standard system debuggers may produce verbose, unstructured outputs (e.g., raw memory dumps) that can quickly exhaust the LLM's context window.
To enable scalable introspection of large data structures, \tool\ employs context compaction strategies that allow the LLM to directly summarize or emit Python scripts that execute against the debugger's output as needed (Line~\ref{algo:summarize_state}).
These scripts, optionally emitted by the LLM (Line~\ref{algo:llm_query}), can extract specific patterns by operating on the raw output data in a sandboxed environment, such as tallying corrupted chunks or traversing linked lists, and return a distilled, semantic summary to the model.
This mechanism preserves critical insights while strictly bounding context usage.
With the support of interactive debugging and context optimization, \tool\ iterates through this introspection loop until the root cause is reliably isolated as recognized by the LLM.
Notably, since LLMs cannot guarantee the precision of their assumptions, \tool\ may exit the debugging loop with a concrete hypothesis that is not fully verified, relying on the patch validation phase to provide empirical feedback for refinement.

\subsection{Patching and Validation}
\label{sec:workflow_phase3}

Once the root cause is isolated, \tool\ transitions from investigation to remediation (Lines~\ref{algo:patch:start}--\ref{algo:patch:end}). 
The agent synthesizes a targeted semantic patch, formatted as a multi-hunk unified diff.
While LLMs are adept at generating corrective logic, they frequently struggle with structural precision, often hallucinating line numbers or context lines~\cite{yu2025patchagent}. 
Because standard patch utilities strictly reject malformed diffs, these superficial formatting errors can trigger premature validation failures, forcing the system into unnecessary and expensive LLM regeneration cycles.
To alleviate this issue, \tool\ applies a deterministic patch correction algorithm~\cite{yu2025patchagent} prior to application.
Instead of relying on the LLM to self-correct trivial alignment errors, this algorithm automatically repairs the diff using a minimal edit distance heuristic. 
It first extracts the unchanged and deleted lines from the proposed patch to reconstruct the intended target code snippet. 
It then slides this snippet across the original source file to find the most probable target range based on edit distance, breaking ties by selecting the location closest to the LLM's original line numbers. Finally, the algorithm updates the diff with the exact contextual lines and correct offsets. This lightweight step ensures that semantically correct patches are not discarded due to minor structural flaws.

Following successful patch application, \tool\ routes this diff to an isolated validation environment. 
The system automatically recompiles the target application and re-executes both the triggering PoC and the test suite $T$. 
If the binary compiles successfully, exits cleanly without triggering sanitizer traps, and passes all functional tests, the patch is marked as verified, and the debugging session terminates successfully (Line~\ref{algo:return_patch}). 
Conversely, if validation fails, \tool\ extracts the resulting compiler errors, sanitizer reports, or test failures. 
To prevent context window exhaustion, the LLM is also utilized to distill these raw logs into concise feedback (Line~\ref{algo:patch:end}) before appending them to the agent's context. 
This execution feedback forces the agent to re-evaluate its prior assumptions, refine its hypothesis, and generate an updated patch. 
This validation loop continues iteratively until a correct patch is synthesized or a predefined iteration limit is reached.

\subsection{Implementation}
\label{sec:implementation}

We implement \tool\ as a decoupled client-server system, utilizing LangChain~\cite{langchain} to manage the core LLM agent workflow.
To support precise codebase navigation, \tool\ integrates a Language Server Protocol (LSP)~\cite{language-server-protocol} front-end with a \texttt{clangd}~\cite{clangd} back-end.
During initialization, a customized compiler wrapper intercepts the project's build process to generate the compilation database required by \texttt{clangd}.
During execution, \tool\ translates LLM-directed codebase queries into standard LSP requests; \texttt{clangd} processes these requests, granting the agent rigorous static analysis capabilities.
For dynamic analysis, \tool\ interfaces the LLM with the debugging environment via the Model Context Protocol (MCP)~\cite{DBLP:journals/corr/abs-2503-23278}.
By standardizing tool invocation and context management through JSON-RPC over standard I/O, MCP functions as an abstraction layer that isolates the non-deterministic reasoning of the LLM from the deterministic, stateful execution of the underlying system tools.
Additionally, we select GDB as the online debugger for its ubiquity and extensive scripting capabilities, while integrating Mozilla \texttt{rr} for deterministic record-and-replay debugging and \texttt{pwndbg} for enhanced memory introspection.

%% file: content/evaluation.tex
\newcommand{\cis}{C}
\newcommand{\nis}{\overline{C}}

This section presents a comprehensive evaluation of the effectiveness of \tool.

\subsection{Evaluation Setup}
\label{subsec:EvaluationSetup}

\mypara{Research Questions.}
The experiments conducted aim to answer the following research questions:

\begin{itemize}[leftmargin=*]

\item \textbf{RQ1:}  Can \tool~repair memory safety vulnerabilities effectively? (Section~\ref{sec:rq1})
\item \textbf{RQ2:} How does \tool\ compare to state-of-the-art LLM-based patching agents in resolving memory safety vulnerabilities? (Section~\ref{sec:rq2})
\item \textbf{RQ3:} How do the components individually and synergistically contribute to the agent's fault localization and patch generation? (Section~\ref{sec:rq3})

\end{itemize}

\mypara{Benchmark.} 
We evaluate \tool\ on SEC-bench~\cite{lee2025secbench}, a comprehensive suite of real-world C/C++ security vulnerabilities. 
SEC-bench is specifically designed for automated program repair tasks where an agent must synthesize a secure patch given an issue report, the target codebase, and a working PoC exploit. 
To ensure reproducibility, each vulnerability is isolated within a Docker container and includes a \texttt{secb} utility that automatically compiles the target and executes the PoC to trigger the corresponding sanitizer error.

We select SEC-bench because it reflects the complexity of production software and facilitates direct comparison with recent state-of-the-art baselines, such as VulnResolver~\cite{DBLP:journals/corr/abs-2601-13933}. 
The dataset encompasses 200 vulnerabilities across 29 diverse open-source projects (e.g., ImageMagick, mruby, gpac) and spans 16 Common Weakness Enumeration (CWE) classes, predominantly memory-safety violations. 
Resolving these vulnerabilities requires deep structural understanding; as noted in prior literature~\cite{DBLP:journals/corr/abs-2601-13933}, the average patch modifies 1.28 files, 2.43 hunks, and 17.29 lines of code, confirming that the benchmark tests well beyond trivial, localized bug fixes.
The benchmark also provides a standardized evaluation and validation environment, enabling a rigorous assessment of \tool's effectiveness in generating correct and secure patches for real-world vulnerabilities.

\mypara{LLM Backbone.}
In this study, we utilize DeepSeek-V3.2~\cite{DBLP:journals/corr/abs-2512-02556} as the default backbone LLM for \tool.
We select DeepSeek-V3.2 because it is an open-source LLM that has demonstrated strong performance on code generation tasks, which are critical for vulnerability resolution.
Particularly, DeepSeek-V3.2 is also utilized in prior work VulnResolver~\cite{DBLP:journals/corr/abs-2601-13933}, allowing for a direct comparison with the baselines evaluated on the same benchmark.
To investigate the generalizability of \tool, we also evaluate its performance with other LLM backbones, including Gemini-3 Flash~\cite{google2025gemini3flash} and GLM-5~\cite{glm5team2026glm5vibecodingagentic}.
Concretely, Gemini-3 Flash is a closed-source, commercial, and cost-efficient LLM that has demonstrated strong performance on code-related tasks, making it a valuable point of comparison to assess the robustness of \tool's design across different LLM architectures and training paradigms.
GLM-5 is one of the most advanced and powerful open-source LLMs available, achieving state-of-the-art performance on various tasks.
By evaluating \tool\ with multiple LLM backbones, we can comprehensively investigate the robustness and generalizability of our approach.

\begin{table}[t]
\centering
\caption{Overall resolution rates and cost efficiency across different LLM backbones on SEC-bench. Costs are reported in USD (\$).}
\label{tab:rq1}
\vspace{1em}
\small
\setlength{\tabcolsep}{0.6em}
\renewcommand{\arraystretch}{1.2}
\begin{tabular}{l|ccc|c}
\toprule
\textbf{Model} & \textbf{Resolved} & \multicolumn{2}{c|}{\textbf{Cost}} & \textbf{Iteration} \\
 & & \textit{Avg.} & \textit{Median} &  \\
\midrule
DeepSeek-V3.2 & 179 (89.5\%) & 0.09 & 0.06 & 18.7 \\
Gemini-3-Flash & 185 (92.5\%) & 0.64 & 0.46 & 24.6 \\
GLM-5 & 189 (94.5\%) & 0.17 & 0.06 & 13.6 \\
\bottomrule
\end{tabular}
\end{table}

\mypara{Configuration.}
\tool's configuration can be specified by users, including the maximum number of iterations, the temperature for LLM generation, and the maximum cost for debugging an issue.
To ensure a fair and consistent evaluation, we employ the same configuration for \tool\ as used in the evaluation of SEC-bench~\cite{lee2025secbench}.
Specifically, we set the maximum number of iterations to 75, the temperature to 0.0 for all LLMs, and the maximum cost to \$1 for GLM-5, DeepSeek-V3.2, and Gemini-3 Flash.
As noted in the original evaluation of SEC-bench~\cite{lee2025secbench}, these settings are chosen to balance the trade-off between performance and cost, ensuring that the evaluation is both rigorous and practical for real-world applications.
Specifically, the low temperature setting encourages deterministic outputs from the LLM, while the iteration and cost limits ensure that the evaluation remains feasible within a reasonable time frame and budget.
By adhering to the same configuration as the original evaluation of SEC-bench, we can directly compare \tool's performance with the baselines evaluated on the same benchmark (\eg\ SWE-agent~\cite{yang2024sweAgent} and OpenHands~\cite{wang2024openhands}), providing a clear assessment of its effectiveness in resolving real-world vulnerabilities.

\mypara{Hardware Configuration.}
All experiments were conducted on an Intel Xeon Gold 6226 24-core processor running at 2.70 GHz with 256 GB of RAM and 2.4 TB of SSD storage.

\begin{figure}
  \centering
  \includegraphics[width=0.7\linewidth]{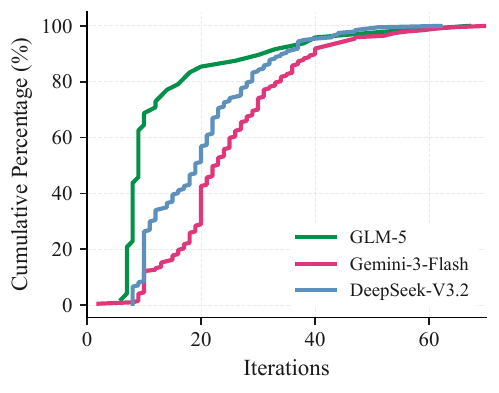}
  \caption{Cumulative distribution of iteration counts for repairs across different LLM backbones. \label{fig:iteration_ecdf}}
\end{figure}

\subsection{RQ1: Effectiveness of \tool}
\label{sec:rq1}

\begin{figure}[t]
\centering
\includegraphics[width=0.55\linewidth]{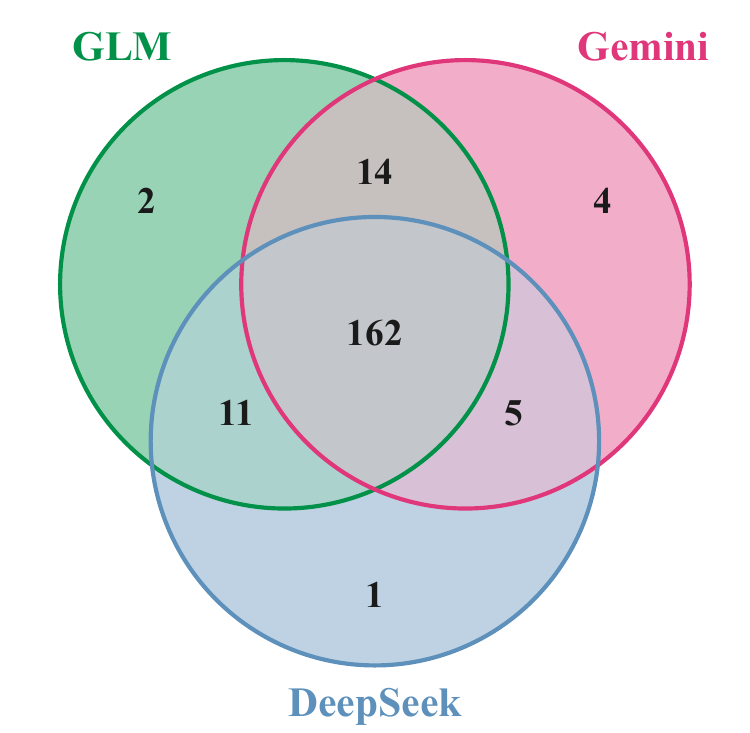}
\caption{Venn diagram showing the overlap of successfully resolved vulnerabilities across the three LLM backbones.\label{fig:venn}}
\end{figure}

To evaluate the effectiveness of \tool\ in repairing memory safety vulnerabilities, we measure its success rate in generating plausible patches across three LLM backbones: GLM-5, DeepSeek-V3.2, and Gemini-3 Flash.
A patch is considered successful if it passes the standardized validation process provided by the SEC-bench framework.
Beyond resolution rates, we analyze iteration patterns, cost efficiency, and patch characteristics to understand how \tool\ leverages dynamic debugging to resolve complex vulnerabilities.
Table~\ref{tab:rq1} summarizes the resolution performance across all backbones.
\tool\ achieves consistently high effectiveness, with resolution rates ranging from 89.5\% to 94.5\%.
Specifically, GLM-5 attains the highest success rate at 94.5\%, followed by Gemini-3 Flash and DeepSeek-V3.2.
The narrow performance spread, only 5 percentage points across models with vastly different training methodologies and scales, indicates that \tool's interactive debugging paradigm effectively amplifies the capabilities of diverse LLM architectures rather than depending on any single model's idiosyncrasies.
Additionally, the Venn diagram in Figure~\ref{fig:venn} illustrates that while there is substantial overlap in the vulnerabilities resolved by all three backbones, each model also uniquely resolves a small subset of cases.
In other words, by combining the three backbones, \tool\ can address 97.5\% of the vulnerabilities in the benchmark, suggesting that the different reasoning styles and strengths of each LLM can complement each other to maximize overall resolution coverage.

In addition, we observe that most (over 70\%) of the vulnerabilities are resolved within 30 iterations, as depicted by the cumulative distribution of iteration counts in Figure~\ref{fig:iteration_ecdf}.
Nonetheless, once the iteration count exceeds 40, the possibility of successful repair sharply declines, indicating that the agent is likely pursuing an incorrect debugging path and exhausting its iteration budget without convergence.
Regarding cost efficiency, the economic cost of operating \tool\ varies across backbones due to differing pricing models and token efficiencies, as reported in Table~\ref{tab:rq1}.
DeepSeek-V3.2 emerges as the most economical option, averaging \$0.09 per vulnerability, while Gemini-3 Flash costs \$0.64 on average.
In total, the average cost per successfully resolved vulnerability remains reasonable across all backbones, especially considering the high resolution rates achieved.

\begin{findingbox}
\textit{\tool~is effective in fixing memory-safety vulnerabilities, achieving 89.5--94.5\% success rates across all LLM backbones, with reasonable cost efficiency.}
\end{findingbox}

\subsection{RQ2: Comparison with Baselines}
\label{sec:rq2}

To assess \tool's effectiveness relative to existing approaches, we compare against state-of-the-art LLM-based repair systems: general-purpose agents SWE-agent~\cite{yang2024sweAgent}, OpenHands~\cite{wang2024openhands}, and Aider~\cite{gauthier2024aider}, as well as specialized vulnerability repair tools PatchAgent~\cite{yu2025patchagent} and VulnResolver~\cite{DBLP:journals/corr/abs-2601-13933}.
To ensure a fair comparison and avoid the prohibitive cost of re-evaluating all baselines with different LLM backbones, we report the baseline results as documented in prior work~\cite{DBLP:journals/corr/abs-2601-13933}.
Their evaluation utilized 80 samples from the SEC-bench dataset, employing DeepSeek-V3.2 as the underlying backbone.
This allows us to directly compare \tool's performance against these baselines under the same conditions, excluding the variable of LLM choice, and focus on the impact of \tool's interactive debugging paradigm on vulnerability resolution effectiveness.

Table~\ref{tab:rq2} presents the issue resolution performance of \tool\ compared to the baselines.
Among specialized tools for vulnerability repair, VulnResolver reaches 67.5\% and PatchAgent achieves 57.5\%, while general-purpose agents lag significantly.
\tool\ achieves the highest resolution rate, outperforming all baselines by a significant margin.
Thus, we deduce that \tool's interactive debugging paradigm provides superior capabilities for localizing and repairing complex memory-safety vulnerabilities, enabling it to generate correct patches for a substantially larger portion of the benchmark compared to existing approaches.
Regarding cost efficiency, \tool\ operates at \$0.09 per vulnerability, which is comparable to VulnResolver (\$0.07) and PatchAgent (\$0.10).
Considering \tool's substantially higher resolution rate, it is more cost-effective, delivering superior value for security-sensitive applications.

\begin{findingbox}
\textit{\tool~outperforms state-of-the-art repair agents, achieving approximately 90\% resolution compared to 67.5\% for the strongest baseline, while maintaining comparable cost efficiency.}
\end{findingbox}

\begin{table}[t]
\centering
\caption{Issue resolution performance on SEC-bench (DeepSeek-V3.2 backbone). For baselines, we report results from prior work~\cite{DBLP:journals/corr/abs-2601-13933}.}
\label{tab:rq2}
\vspace{1em}
\small
\setlength{\tabcolsep}{0.8em}
\renewcommand{\arraystretch}{1.1}
\begin{tabular}{l|c|c}
\toprule
\textbf{Technique} & \textbf{Resolved (\%)} & \textbf{Avg. Cost (\$)} \\
\midrule
\multicolumn{3}{l}{\textit{General-Purpose Agents}} \\
\midrule
SWE-agent~\cite{yang2024sweAgent} & 37.5 & 0.11 \\
OpenHands~\cite{wang2024openhands} & 20.0 & 0.05 \\
Aider~\cite{gauthier2024aider} & 20.0 & 0.04 \\
\midrule
\multicolumn{3}{l}{\textit{Specialized Vulnerability Repair}} \\
\midrule
PatchAgent~\cite{yu2025patchagent} & 57.5 & 0.10 \\
VulnResolver~\cite{DBLP:journals/corr/abs-2601-13933} & 67.5 & 0.07 \\
\midrule
\midrule
\textbf{\tool}  & \textbf{89.5} & 0.09 \\
\bottomrule
\end{tabular}
\vspace{-1em}
\end{table}

\subsection{RQ3: Ablation Study}
\label{sec:rq3}

To understand the contribution of each component to \tool's effectiveness, we conduct an ablation study using DeepSeek-V3.2 as the backbone.
Specifically, we disable key debugging capabilities and measure the impact on resolution rate.
Concretely, we evaluate the following two variants:

\begin{itemize}[leftmargin=*]
  \item \textbf{GDB-only}: This variant removes support for \texttt{rr}'s deterministic replay and \texttt{pwndbg}'s heap introspection, leaving only standard GDB debugging capabilities.
  \item \textbf{w/o debugger}: This variant removes all dynamic debugging support, forcing the agent to rely solely on static information under the guidance of our signature-driven prompting strategy.
\end{itemize}

Table~\ref{tab:rq3} summarizes the resolution performance of each variant.
Specifically, both the \texttt{GDB-only} and \texttt{w/o debugger} variants show degradation in terms of resolution rate, with the latter exhibiting a more pronounced drop, achieving 82.0\% and 77.0\% respectively, compared to 89.5\% for the full configuration.
Moreover, since our approach is designed based on PatchAgent~\cite{yu2025patchagent}, which does not utilize signature-driven initialization and interactive state introspection, we can understand the contribution of the signature-driven initialization by comparing the \texttt{w/o debugger} variant with PatchAgent.
We observe that the \texttt{w/o debugger} variant still achieves a higher resolution rate than PatchAgent and other baselines, indicating that our signature-driven prompting strategy provides a strong foundation for vulnerability repair even without dynamic debugging support.
In addition, regarding the iteration count required for successful repairs, we find that there is no significant difference between the full configuration and the variants, suggesting that the iteration count appears to be more influenced by the LLMs' reasoning process rather than the presence of interactive debugging.

\begin{figure}[t]
  \centering
  \includegraphics[width=0.7\linewidth]{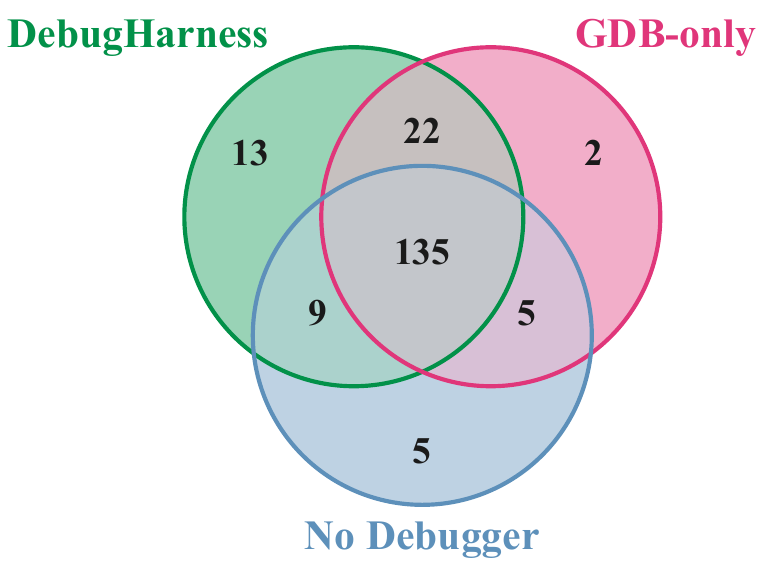}
  \caption{Venn diagram showing the overlap of successfully resolved vulnerabilities across the full configuration and the two ablation variants. \label{fig:ablation_venn}}
\end{figure}

To investigate the statistical significance of these differences, we perform Pearson's chi-squared ($\chi^2$) test~\cite{Pearson1992} to compare the resolution rates of each variant against the full configuration.
We employ Pearson's chi-squared test because it is appropriate for comparing proportions across independent groups with categorical outcomes, which in this case is whether each vulnerability was resolved or not.
The test evaluates whether the observed difference in resolution rates between each variant and the full configuration could arise by chance alone.
As depicted in Table~\ref{tab:rq3}, these differences are statistically significant, with the \texttt{GDB-only} variant showing a significant degradation ($\chi^2 = 4.60$, $p < 0.05$) and the \texttt{w/o debugger} variant exhibiting an even more significant drop ($\chi^2 = 11.21$, $p < 0.001$).
However, while \tool\ and the \texttt{GDB-only} variant both achieve higher resolution rates than the \texttt{w/o debugger} variant, it does not mean that all the vulnerabilities resolved by the \texttt{w/o debugger} variant can also be resolved by \tool\ and the other variant.
Concretely, as shown in Figure~\ref{fig:ablation_venn}, there are 5 vulnerabilities that can only be resolved by the \texttt{w/o debugger} variant.
Through manual analysis, we find that these cases mostly involve compiler optimization, which makes the debugging information inaccurate or even unavailable, thus potentially misleading \tool\ to generate incorrect patches.

\begin{table}[t]
\centering
\caption{Ablation study results: contribution of each component (DeepSeek-V3.2 backbone). $\Downarrow$ indicates the percentage point drop in resolution rate compared to the full configuration. Statistical significance is assessed via Pearson's chi-squared test.}
\label{tab:rq3}
\vspace{1em}
\small
\renewcommand{\arraystretch}{1.1}
\begin{tabular}{l|ccc}
\toprule
\textbf{Configuration} & \textbf{Resolved} & \textbf{Degradation} & \textbf{$\chi^2$} \\
\midrule
\tool & 179 (89.5\%) & -- & -- \\
\midrule
\enspace -- GDB-only & 164 (82.0\%) & $\Downarrow$ 7.5\% & $4.60^*$ \\
\enspace -- w/o debugger & 154 (77.0\%) & $\Downarrow$ 12.5\% & $11.21^{***}$ \\
\bottomrule
\end{tabular}
\footnotesize $^* p < 0.05, ^{***} p < 0.001$ via Chi-squared test.
\vspace{-0.5em}
\end{table}

\mypara{Vulnerability Type Analysis.}
To further understand the advantage provided by the dynamic debugging capabilities, we break down the resolution rates by vulnerability type across the ablation variants.
Specifically, we analyze the resolution rates for the most common vulnerability types in the benchmark, including Heap Use-After-Free (UAF), Null Pointer Dereference, Heap Buffer Overflow, Memory Leak, and Stack Buffer Overflow.
Among them, Heap UAF suffers the largest degradation when advanced debugging features are removed, dropping from 86.7\% to 70.6\%, indicating that tracing object lifecycles through reverse execution is essential for resolving temporal memory bugs.
Other vulnerability types, such as Heap Buffer Overflow (92.9\% $\rightarrow$ 84.7\%) and Memory Leak (66.7\% $\rightarrow$ 53.3\%), also degrade notably without heap introspection to identify bounds violations and track allocation sites.
For Null Pointer Dereference, repairs degrade most significantly (90.8\% $\rightarrow$ 70.8\%), a 20 percentage point drop indicating that locating null sources requires iterative state inspection beyond static analysis.
Interestingly, Stack Buffer Overflow remains resilient, suggesting these localized spatial bugs can also be fixed from sanitizer reports alone.
In conclusion, the ablation study confirms that dynamic debugging significantly contributes to \tool's superior performance, especially for temporal vulnerabilities.

\begin{findingbox}
\textit{The ablation study confirms that the components of \tool contribute synergistically to its superior performance, with dynamic debugging capabilities providing significant advantages for localizing and repairing complex memory-safety vulnerabilities.}
\end{findingbox}

%% file: content/discussion.tex
In this section, we further discuss the threats to validity, limitations, and future work of our investigation.

\subsection{Threats to Validity}
Our investigation faces several threats to validity that merit attention. 
First, configuration choices, such as the selection of temperature and maximum iterations, may influence the performance of \tool.
To mitigate this, we adopt the same configuration as the original evaluation of SEC-bench~\cite{lee2025secbench}, ensuring a consistent basis for comparison with existing baselines.
Second, the choice of LLM backbone also presents a potential threat to validity, as different models may exhibit varying capabilities in code generation and reasoning tasks.
To address this, we evaluate \tool\ with multiple LLM backbones, including GLM-5, DeepSeek-V3.2, and Gemini-3 Flash, to assess the robustness of our approach across different model architectures and training paradigms.
We observe that \tool\ consistently demonstrates strong performance across all evaluated LLMs, suggesting that its effectiveness is not solely dependent on a specific model choice.
Lastly, data contamination is also a concern, as the presence of test cases in the training data can artificially inflate performance metrics.
To mitigate this concern, we directly compare \tool's performance with baselines~\cite{DBLP:journals/corr/abs-2601-13933} powered by the same LLM backbones on the same benchmark, demonstrating that the observed improvements are attributable to \tool's design rather than differences in training data.

\subsection{Limitations}

While \tool\ significantly advances automated vulnerability resolution through dynamic introspection, its reliance on runtime execution introduces specific structural and operational constraints. 
As illustrated in Section~\ref{sec:rq3}, a primary limitation arises when diagnosing vulnerabilities that only manifest under aggressive compiler optimizations (e.g., \texttt{-O2} or \texttt{-O3}). 
In such cases, optimizations such as instruction reordering, function inlining, and aggressive register allocation obscure the mapping between the executing binary and the original source code. 
Consequently, interactive debuggers frequently report variables as optimized out or map execution states to imprecise line numbers. 
Because \tool\ fundamentally depends on accurate, human-readable dynamic feedback to guide the LLM's reasoning, these optimization-induced distortions can mislead the agent, potentially resulting in hallucinated root causes or failed patch generation. 
Extending the harness to natively reason over optimized assembly or intermediate representations remains an open challenge.
In practical scenarios, \tool\ can leverage heuristic strategies to mitigate this issue, such as automatically adjusting compiler flags to reduce specific optimizations interfering with debugging information, or selectively disabling debugging support for certain code regions known to be heavily optimized.
These strategies can alleviate the negative impact of compiler optimizations on \tool's performance.

Additionally, while \tool\ has demonstrated strong performance in resolving vulnerabilities, the incorporation of dynamic debugging introduces additional overhead compared to purely static approaches.
This overhead may limit the applicability of \tool\ in scenarios where rapid patch generation is required.
According to our internal measurements, the average time taken for the harness to interact with the debugging tools accounts for 11\% of the total resolution time, which is manageable considering the significant improvement in resolution rate.
However, further optimizations to reduce this overhead, such as caching intermediate debugging results or parallelizing certain diagnostic steps, could enhance \tool's efficiency and make it more suitable for time-sensitive applications.

\subsection{Future Work}

\tool's current design and implementation are focused on resolving memory safety vulnerabilities in C/C++ applications, leveraging sanitizers and language servers to guide the dynamic debugging process.
In principle, the idea of \tool\ is not limited to this specific class of bugs.
For instance, interactive debuggers are regarded as the most powerful tools for diagnosing complex concurrency bugs, such as data races, which are challenging for existing LLM-based agents to resolve.
Hence, we believe that \tool\ has the potential to be extended to resolve concurrency bugs by leveraging the capabilities of interactive debuggers---a critical area for improving software reliability and security in practice.
We plan to explore this aspect in future work, extending \tool's applicability to a broader spectrum of software issues, including logic bugs and performance bottlenecks.
In addition, \tool\ is designed to be modular and extensible, allowing for the seamless integration of additional debugging tools to enhance its diagnostic capabilities.
Future work will explore the integration of supplementary diagnostic utilities, such as Valgrind~\cite{DBLP:conf/pldi/NethercoteS07} for memory profiling or \texttt{strace}~\cite{strace} for system call tracing, which can provide deeper insights into program behavior and facilitate the diagnosis of a wider range of bugs.
This integration can be achieved by deploying these tools as independent MCP servers, without modifying the core harness logic, thus preserving system modularity.
Our evaluation of \tool\ currently focuses on vulnerability benchmarks in C/C++ applications; we plan to extend its application to more complex systems, such as the Linux kernel, and projects in other programming languages.
Investigating \tool's performance on real, unresolved bugs in open-source projects and proposing effective patches via Pull Requests will be an interesting future direction as well.

%% file: content/relatedworks.tex
In this section, we situate \tool\ within the broader landscape of automated software maintenance, tracing the evolution from general Automated Program Repair (APR) to specialized Automated Vulnerability Repair (AVR). 

\subsection{Software Issue Resolution}

Automated Program Repair (APR) encompasses a wide range of techniques designed to autonomously resolve software defects~\cite{liu2019tbar, le2012GenProg}.  
In recent years, the advent of LLMs has fundamentally transformed this domain, enabling systems to reason about complex codebase semantics~\cite{xia2024chatRepair, yin2024thinkrepair, bouzenia2024repairagent, zhang2025ReinFix}.
This progress has been significantly accelerated by realistic benchmarks such as SWE-bench~\cite{jimenez2023swebench}, which evaluates agents on real-world GitHub issues. 

Modern LLM-based APR systems generally fall into two paradigms: agent-based and workflow-based. 
Agent-based frameworks, such as SWE-agent~\cite{yang2024sweAgent} and OpenHands~\cite{wang2024openhands}, equip LLMs with tools (e.g., language servers) to iteratively explore repositories and synthesize patches. 
By contrast, workflow-based systems like Agentless~\cite{xia2025agentless} decompose issue resolution into rigid, sequential phases, such as fault localization and patch generation, achieving competitive accuracy without the overhead of open-ended tool use. 
Recent efforts have further expanded these paradigms to specialized domains, including multimodal GUI repair~\cite{huang2025guirepair, yang2024sweBenchMultimodal} and meta-agent debugging~\cite{rahardja2025canAgentsFixAgentIssues}. 
While these general-purpose APR systems excel at resolving bugs, especially issues in Python projects as collected in SWE-bench, they struggle to resolve bugs in C/C++ projects~\cite{DBLP:journals/corr/abs-2504-02605}, particularly memory-safety vulnerabilities, as indicated by prior studies~\cite{lee2025secbench, DBLP:journals/corr/abs-2601-13933}.
Therefore, we propose \tool\, aiming to resolve vulnerabilities in C/C++ applications. 
It incorporates interactive state introspection, which is essential for diagnosing and repairing complex memory-safety issues, thus significantly advancing the capabilities of LLM-based agentic systems in this critical domain.

\subsection{Automated Vulnerability Repair}

As a specialized subset of APR, Automated Vulnerability Repair (AVR) targets security flaws~\cite{li2025sokAVR}. 
These issues attract significant attention due to their potentially critical impact on system security~\cite{DBLP:journals/corr/abs-2511-11019}.
Early LLM-based AVR approaches, such as VulRepair~\cite{fu2022vulrepair}, VRepair~\cite{chen2023vrepair}, and APPATCH~\cite{nong2025appatch}, demonstrated the viability of neural patch generation but heavily relied on human-provided fault locations or predefined CWE labels. 
More recent systems attempt to fully automate the AVR pipeline.
For instance, PatchAgent~\cite{yu2025patchagent} and SAN2PATCH~\cite{kim2025SAN2PATCH} utilize sanitizer logs to guide fault localization. 
However, these systems treat crash logs as static text queries for context retrieval, mirroring the behavior of general-purpose APR agents like MarsCoder~\cite{liu2024marscode}.
By failing to exploit the deep runtime semantics embedded in the execution state, these approaches struggle to differentiate themselves from standard APR frameworks and often fall short when diagnosing complex memory corruptions. 

An alternative trajectory in AVR focuses on safety properties. 
Pre-LLM systems like Senx~\cite{useSafePropToGenVulPatch} employed symbolic execution to mine and enforce safety constraints. 
While precise, symbolic execution is fundamentally bottlenecked by path explosion and is typically restricted to predefined bug templates (e.g., simple buffer overflows). 
To bypass these scalability issues, recent LLM-driven systems like VulnResolver~\cite{DBLP:journals/corr/abs-2601-13933} attempt to autonomously mine and validate safety properties, offering deeper semantic insights than simple context retrieval. 
Yet, even these property-centric approaches remain largely static in their reasoning, lacking the ability to introspect the execution state of the vulnerable program.

In practice, human security experts do not resolve complex vulnerabilities by merely reading source code and crash logs. Instead, they rely on interactive debuggers to step through execution, inspect memory layouts, and observe state mutations chronologically. 
\tool\ is explicitly designed to emulate this expert workflow. 
Unlike existing AVR methods that rely on static context retrieval or symbolic property generation, \tool\ bridges static code analysis with dynamic state introspection.
By integrating signature-driven initialization with interactive debugging capabilities, \tool\ empowers the LLM agent to systematically test root-cause hypotheses against the actual runtime state.
Our evaluation on SEC-bench~\cite{lee2025secbench} demonstrates that this dynamic orchestration allows \tool\ to significantly outperform state-of-the-art baselines, including both general APR agents and recent vulnerability-specific tools like PatchAgent and VulnResolver.

%% file: content/conclusion.tex
In this paper, we presented \tool, an autonomous agent harness that fundamentally shifts the paradigm of automated vulnerability repair from static guessing to dynamic, evidence-based debugging.
By emulating the investigative workflows of human security analysts, \tool\ actively engages with the failing program's execution state.
\tool\ leverages signature-driven initialization to guide the LLM's search using domain-specific heuristics.
Additionally, interactive state introspection equips the harness with interactive debugging and deep memory analysis capabilities. 
Finally, our closed-loop patching and validation pipeline enables the agent to iteratively refine its hypothesis and patch based on empirical feedback.
Our evaluation on the SEC-bench dataset confirms the efficacy of our approach. 
\tool\ significantly outperforms state-of-the-art general APR agents and specialized vulnerability repair systems, and ablation studies confirm that its components contribute to this superior performance.